\documentclass[10pt]{article}
\usepackage[utf8]{inputenc}                                                     
\usepackage[margin=0.5in]{geometry}                                             
\usepackage{amsmath}                                                            
\usepackage{amsfonts}                                                           
\usepackage{graphics}
\usepackage{cite}
\usepackage{url}
\usepackage{booktabs}
\usepackage{graphicx}
\usepackage{subcaption}
\usepackage[usenames, dvipsnames]{color}
\usepackage[linesnumbered]{algorithm2e}

\long\def\ca#1\cb{} 

\newcommand{\becs}{\begin{cases}}
\newcommand{\bem}{\begin{matrix}}

\newcommand{\bra}[1]{\langle#1|}

\newcommand{\bsk}{\bigskip }

\newcommand{\encs}{\end{cases}}
\newcommand{\enm}{\end{matrix}}

\newcommand{\inpV}[2]{\langle#1,#2\rangle }
\newcommand{\ket}[1]{|#1\rangle }

\newcommand{\lgl}{\langle } 

\newcommand{\ot}{\otimes }

\newcommand{\rgl}{\rangle }

\newcommand{\Tr}{{\rm Tr}}

\newcommand{\AC}{{\mathcal A}}
\newcommand{\BC}{{\mathcal B}}

\newcommand{\DC}{{\mathcal D}}

\newcommand{\FC}{{\mathcal F}}

\newcommand{\HC}{{\mathcal H}}
\newcommand{\IC}{{\mathcal I}}

\newcommand{\LC}{{\mathcal L}}
\newcommand{\MC}{{\mathcal M}}
\newcommand{\NC}{{\mathcal N}}

\newcommand{\SC}{{\mathcal S}}
\newcommand{\TC}{{\mathcal T}}

\newcommand{\dB}{\textbf{d}}

\newcommand{\rB}{\textbf{r}}

\newcommand{\xB}{\textbf{x}}
\newcommand{\yB}{\textbf{y}}

\newcommand{\bt}{\beta }

\newcommand{\dl}{\delta }

\newcommand{\lm}{\lambda }
\newcommand{\Lm}{\Lambda }

\newcommand{\sg}{\sigma }

\newcommand{\Up}{\Upsilon }

 \def\outl#1{}  \def\xa{} \def\xb{}  
\begin{document}

\begin{center}
{\bf Maximum a posteriori probability estimates for quantum tomography}\bsk\\
\normalsize Vikesh Siddhu \\
\textit{Department of Physics, Carnegie Mellon University, Pittsburgh, 
  Pennsylvania 15213, U.S.A.}\\
Date: 28$^{\text{th}}$ Jan'19
\vspace{.1cm}
\end{center}

 \begin{abstract}
\xb \outl{MAP estimation, importance of including prior, BME and computational difficulty,
connection to other estimators, noisy measurement, qubit numerics, algorithm
for optimization} \xa

Using a Bayesian methodology, we introduce the maximum a posteriori probability~(MAP)
estimator for quantum state and process tomography.
We show that the maximum likelihood, the hedged maximum likelihood,
and the maximum likelihood-maximum entropy estimator, and estimators of this general type,
can be viewed as special cases of the MAP estimator. 
The MAP, like the Bayes’ mean estimator includes prior knowledge.
For cases of interest to tomography MAP can take advantage of 
convex optimization tools making it numerically tractable.
We show how the MAP and other Bayesian quantum state estimators can be corrected for 
noise produced if the quantum state passes through a noisy quantum channel prior to measurement.
Numerical simulations on a single qubit indicate that on average,
including these corrections significantly improve the estimate 
even when the measurement data are modestly large.

\end{abstract}
\section{Introduction}
\label{sct1} 
\xb \outl{Tomography, point and interval estimate, including prior knowledge, 
Bayes’ Mean Estimator and drawback, efficiently computable MAP, results for Numerical simulation 
under noise, paper organization} \xa
The task of quantum state tomography is to estimate a quantum state 
or density operator by performing measurements.
Its classical analogue is to estimate the parameters of a probability distribution by 
sampling from it several times. Quantum process tomography deals with the estimation 
of a noisy quantum channel or completely positive trace preserving map;
its classical analogue is the estimation of a conditional
probability distribution.

An \textit{estimator} is a procedure which uses the data from measurements to
construct an \textit{estimate} of the object of interest
called the \textit{estimand}. 
A \textit{point estimator} provides 
a single best guess of the estimand, for example guessing 
the bias of a coin by flipping it several times, or locating a point
in the qubit Bloch ball by measuring many identically prepared qubits. An 
\textit{interval estimator}, more generally a \textit{set estimator}, provides
a set of plausible values for the estimand, 
for example a confidence interval for the bias of a coin, 
or a confidence region in the Bloch ball for a qubit state. 
In general, interval estimates provide more information, such as error
bars, but are harder to construct than point estimates. For the latter error
bars must be constructed independently.
A lot of effort has been devoted to constructing good estimators for 
quantum states. Various point estimators~\cite{PhysRevA.55.R1561,
PhysRevLett.105.200504, PhysRevLett.107.020404, PhysRevLett.105.150401,teoBook}
and interval estimators~\cite{Blume-Kohout2012, Ferrie2016, Christandl2012,PhysRevLett.117.010404}
have been proposed. 

Maximum A Posteriori~(MAP) point estimators are widely used in
statistics, and have been applied in various fields of 
physics~\cite{doi:10.1117/12.2212329, Gursoy2015}.
In this work we introduce the MAP estimator for quantum state and channel tomography.
One begins with a prior probability density on the set of quantum states or channels, 
and using the measurement data the prior is updated to obtain a posterior density. 
The maximum of the posterior density gives the MAP estimate. The mean of the posterior
density gives what is called the Bayes’ mean estimator~(BME)~\cite{Blume-Kohout2010}.
We show that in many cases of interest in quantum tomography
the MAP estimator can be computed efficiently using convex optimization 
tools~\cite{boydVan04}, some of these tools have been tailored for
quantum information~\cite{PhysRevA.95.062336, Bolduc2017, arXiv:1803.10062}.
Obtaining the MAP estimate may be computationally simpler than computing the BME, 
which is evaluated by numerically integrating over the set of density operators.

Several well known estimators, in particular the maximum likelihood estimator~(MLE)~\cite{PhysRevA.55.R1561},
the hedged maximum likelihood estimator~(HMLE)~\cite{PhysRevLett.105.200504},
and the maximum likelihood-maximum entropy~(MLME) estimator~\cite{PhysRevLett.107.020404},
can be viewed as special cases of the MAP estimator corresponding to particular
choices of the prior probability density. Thus, the MAP estimator
provides a systematic framework for discussing estimators of this general type, and casts them
in a new light. In addition we show how MAP 
estimators can be applied to quantum process tomography.

Experimental setups are noisy, and it is useful to be able to correct the 
experimental data for noise. The MAP approach provides an easy way to do this
if the noise can be represented by a noisy quantum channel with known 
parameters~(that may have been determined in a separate experiment), 
through which the system of interest passes on its way to the measurement device.

The rest of this paper is organised as follows. Section 2 is devoted
to a general discussion of quantum state tomography and
the linear inversion estimator; the material here is not new, but 
helps understand the later material. Section 3 introduces Bayesian estimators and the MAP estimator for
quantum state and process tomography. 
In Sec. 4 we discuss how the simple noise model mentioned above
can be incorporated in the MAP estimation framework, and present
results from simulations that evaluate the effect of incorporating
the noise model. 
Section 5 is a brief summary, and is followed by an
appendix illustrating the use of convex optimization tools for minimizing
convex functions over qubit density operators.

\section{Quantum State Tomography}
\label{sct2} 
Let $\HC$ be a $d$-dimensional Hilbert space, and $\LC(\HC)$ be
the space of operators on $\HC$. The set $\SC$ of quantum states i.e.
density operators forms a convex subset of $\LC(\HC)$. 
Measurements on quantum systems can be described using a 
POVM~(positive operator-valued measure), a collection of
positive operators that sum to the identity in $\HC$.
Let $\{ \Lm_i \}$ be a POVM, and $\rho$ be a density operator.
The probability $p_i$ of observing an outcome corresponding 
to the operator $\Lm_i$ is
\begin{equation}
p_i = \Tr(\rho \Lm_i).
\label{POVMProb}
\end{equation}
One simple measurement scheme for doing quantum state tomography is to prepare $N$
quantum systems, each corresponding to density operator $\rho$, and independently measure 
each using the same POVM $\{\Lm_i\}_{i=1}^k$.
The measurements yield a data set $\dl = \{n_1, n_2,\dots ,n_k \}$, where $n_i$
is the number of measurement outcomes corresponding to $\Lm_i$, and 
$\sum_i n_i = N$. The probability $\Pr(\dl|\rho)$ of observing the data 
set $\dl$ given the density operator $\rho$ is
\begin{equation}
\Pr(\dl|\rho)  = C_{\dl} \overset{k}{\underset{i = 1}{\prod}} p_i^{n_i},
 \label{PdataRho}
\end{equation}
where $C_{\dl} = N/ (n_1!) \dots (n_k!)$ is a normalization constant, which
depends only on $\dl$.

\xb \outl{Linear Inversion, POVM, negative eigenvalue, pInv} \xa
The \textit{linear inversion estimator} $\hat{\rho}_{\text{inv}}$ is a simple method for estimating 
a system's density operator $\rho$  when the measurement data are
related to $\rho$ by a set of invertible linear equations.
An example of this general strategy is
the measurement scheme discussed above when the POVM $\{\Lm_i\}_{i=1}^{d^2}$
forms a basis of $\LC(\HC)$. The dual basis $\{\bar \Lm_i\}$ 
is defined by
\begin{equation}
 \lgl \bar \Lm_i, \Lm_j \rgl = \dl_{ij}, \quad i,j \in \{1, \dots, d^2\},
 \label{dual}
\end{equation}
where $\lgl \rho,\sigma \rgl = \Tr(\rho^{\dag} \sigma)$ is the Frobenius inner product.
The set of invertible linear equations,
\begin{equation}
\Tr(\hat{\rho}_{\text{inv}} \Lm_i) = \hat p_i := n_i/N, \quad i = 1, \dots, d^2,
\label{linInvEq}
\end{equation}
where $\hat p_i$ is an estimate of $p_i$,  
can be solved to obtain the linear inversion estimate
\begin{equation}
 \hat{\rho}_{\text{inv}} = \sum_{i=1}^{d^2} \hat p_i \bar \Lm_i.
 \label{linINV}
\end{equation}
The estimate $\hat p_i$ has a variance of $p_i(1-p_i)/N$, so one
expects this strategy to work well when $N$ is large.
The linear inversion estimator generalises in an obvious way when 
$\{\Lm_i\}$ is a Hermitian basis of the operator space
but not a POVM.

The linear inversion estimator is quite special as it requires a set of invertible
equations of the form \eqref{linInvEq}. Note that estimates constructed using \eqref{linINV}
may not be valid quantum states. While they have unit trace, they may have negative eigenvalues.
\section{Bayesian Estimators}
\label{sct3.1}
\xb \outl{Definition of MAP estimate} \xa
\subsection{Bayes’ Rule}
For Bayesian estimators one chooses a prior probability for the estimand, and
a model that relates observed data to the estimand. Using Bayes’ rule~(discussed below),
the prior is updated to obtain a posterior probability, and the latter is used to construct 
point or set estimates. We will be focusing on point estimates.

Let $\Pr(\rho)$ be a \textit{prior} probability measure on $\SC$. It represents the belief
or uncertainty about the quantum system prior to the measurement. For quantum state tomography,
measurements are performed on many copies of a quantum system to generate a discrete data set $\dl$. 
A model $\Pr(\dl|\rho)$ is chosen, it relates $\dl$ to 
$\rho$~(see Eq. \eqref{PdataRho} for an example) and represents the probability
of obtaining the data given the quantum state. In the literature,
$\Pr(\dl|\rho)$ for a fixed $\dl$ is often viewed as a non-negative function of $\rho$ 
called the \textit{likelihood function}~(for more examples see \cite{Smolin2012, Singh2016}).
The \textit{posterior} probability measure $\Pr(\rho|\dl)$
on $\SC$ given the data are obtained from Bayes’ rule
\begin{equation}
 \Pr(\rho|\dl) = k \Pr(\dl| \rho) \Pr(\rho),
\label{bayes}
\end{equation}
where $k$, which depends on $\dl$, not $\rho$, is a normalization constant.

As $\rho$ varies continuously, a useful way to express $\Pr(\rho)$ is by
making $\rho$ a smooth one to one function $\rho(\xB)$ of a collection
of real variables $X$, and introducing a non-negative prior probability
density $p(\rho)$ such that
\begin{equation}
 \Pr(\rho \in \AC) = \int_{\mathit{A}} p(\rho(\mathbf{x})) d\mathbf{x},
 \label{PDPrior}
\end{equation}
where $\AC \subseteq \SC$ is some subset of density operators, and
$\mathit{A} \subseteq X$ the corresponding subset of parameters.
The posterior probability density
\begin{equation}
 p(\rho|\dl) = k \Pr(\dl|\rho)p(\rho),
 \label{postDen}
\end{equation}
uses the same parametrization $X$ as before, and is
related to the measure $\Pr(\rho|\dl)$ in a manner similar to
\eqref{PDPrior}.
Note that
for a given probability measure $\Pr(\rho)$ the density $p(\rho)$
depends on the choice of parametrization $\rho(\xB)$. Conversely, if $p(\rho)$
is held fixed, a different parametrization will lead to a different measure $\Pr(\rho)$.
The same is true for the relationship between $\Pr(\rho|\dl)$ and $p(\rho|\dl)$.

The Bayes’ mean estimator $\hat\rho_{\text{BME}}$~\cite{Blume-Kohout2010}
is the expectation of $\rho$ in the posterior probability measure $\Pr(\rho|\dl)$, and
can be written using the density $p(\rho|\dl)$ as
\begin{equation}
 \hat\rho_{\text{BME}} = \int \rho(\mathbf{x}) p(\rho(\mathbf{x})|\dl) d\mathbf{x}.
\label{BME}
\end{equation}
While the terms in the integrand depend on the parametrization $X$ used for 
$\rho(\xB)$, the integral itself is independent of the parametrization,
if $\Pr(\rho|\dl)$ is held fixed. If instead $p(\rho|\dl)$ is
fixed, changing the parametrization may change 
$\Pr(\rho|\dl)$~(see comments following \eqref{postDen}) and alter $\hat\rho_{\text{BME}}$.
Evaluating the integral \eqref{BME}
numerically can be cumbersome as for $n$ qubits $X$ consists of 
$2^{2n}-1$ variables.

\subsection{MAP estimate}
\xb \outl{Definitions of MLE, HMLE and MLME, Explicit connection and new perspective} \xa
\label{sct3.2}
The maximum a posteriori probability~(MAP) estimate $\hat \rho_{\text{MAP}}$,
is the density operator $\rho$ for which the posterior probability density
is maximum. Its advantage is that in many cases it can be easily computed.
When the data set is large, one expects for a suitable prior that
$\hat \rho_{\text{BME}}$ and $\hat \rho_{\text{MAP}}$ are close.

Maximizing $p(\rho|\dl)$ is equivalent to maximizing $\log p(\rho|\dl)$,
and since $k$ is independent of $\rho$ it follows from \eqref{postDen} that
\begin{equation}
 \hat{\rho}_{\text{MAP}} = \underset{\rho \in \SC}{\text{argmax}} \; [ \log \Pr(\dl|\rho) + \log p(\rho)].
  \label{MAPForm}
\end{equation}
Notice that for a given $\Pr(\rho)$, and thus $\Pr(\rho|\dl)$ as given
by \eqref{bayes}, $\hat{\rho}_{\text{MAP}}$ will depend on the parametrization
$X$ for $\rho(\xB)$. See the comments above in connection with \eqref{PDPrior}.
Conversely, if $p(\rho)$, and thus $p(\rho|\dl)$ as given by \eqref{postDen}, is held fixed,
changing the parametrization may change $\Pr(\rho)$ and $\Pr(\rho|\dl)$ without
altering  $\hat{\rho}_{\text{MAP}}$.

It is often the case in quantum tomography that 
$\log \Pr(\dl| \rho)$ is concave in $\rho$~(see Eq. \eqref{PdataRho} for an example).
If in addition, as is the case for a number of 
priors~(see below), $\log p(\rho)$ is a concave function of $\rho$
then the same is true for the objective function on the right hand side of
\eqref{MAPForm}, and $\hat{\rho}_{\text{MAP}}$ can be efficiently computed using tools of 
convex optimization. 

If one knows that $\rho$ belongs to a discrete set of possibilities, 
the above discussion is modified in an obvious way; 
$\hat \rho_{\text{BME}}$ is the weighted average of finitely many density operators
computed with respect to the left side of \eqref{bayes}, and $\hat \rho_{\text{MAP}}$
is the density operator for which the left side of \eqref{bayes} is maximum.

The MLE, HMLE, and MLME estimators 
can be viewed as MAP estimators using suitable prior probability densities.
The maximum likelihood estimate~(MLE)
\begin{equation}
 \hat{\rho}_{\text{MLE}} = \underset{\rho \in \SC}{\text{argmax}} \; \log \Pr(\dl|\rho),
  \label{MLE}
\end{equation}
coincides with $\hat{\rho}_{\text{MAP}}$ in \eqref{MAPForm} when the 
prior probability density $p(\rho)$ is independent of $\rho$. 
While MLE and the MAP estimate are the same for this special choice of prior, 
note that: the former is the density operator for which the data are most likely,
and the latter is the most probable density operator given the data and the prior probability density.

The hedged maximum likelihood estimate~(HMLE)
\begin{equation}
 \hat{\rho}_{\text{HMLE}} = \underset{\rho \in \SC}{\text{argmax}} \; \{ \log \Pr(\dl|\rho) + \log [\det(\rho)]^{\bt} \},
 \quad \bt > 0,
  \label{HMLE}
\end{equation}
is of the MAP form, where the prior $p(\rho) \propto [\det(\rho)]^{\bt}$ is 
called the \textit{hedging function}; it guarantees a full rank estimate.
When $\bt$ is an integer, this prior probability density can 
be viewed as a special case of an \textit{induced measure}~(see Eq.~(3.5) in \cite{KZHS01})
obtained by choosing an ancillary system of dimension $k = \bt + d$, defining the 
Haar measure on a $d \times k$ dimensional Hilbert space, then tracing out the 
ancillary system to induce a distribution on the space of $d$ dimensional
density operators. The function $\log [\det(\rho)]^{\bt}$ is concave in $\rho$ for $\bt > 0$,
and when $\log \Pr(\dl|\rho)$ is concave the HMLE can be efficiently computed. 

The maximum likelihood maximum entropy~(MLME) estimate
\begin{equation}
  \hat{\rho}_{\text{MLME}} = \underset{\rho \in \mathbf \SC}{\text{argmax}} \; [ \log \Pr(\dl|\rho) + \lm S(\rho) ], \quad \lm \geq 0,
   \label{MLME}
\end{equation}
is a MAP estimate with a prior which is exponential
in the von-Neumann entropy $S(\rho)=-\Tr(\rho \log \rho)$.
Since $S(\rho)$ is concave in $\rho$, when $\Pr(\dl|\rho)$
is concave the MLME can be efficiently computed. Other possible
advantages of the MLME estimator have been discussed in~\cite{PhysRevLett.107.020404}.

\subsection{MAP and quantum process tomography}
\label{sct3.3}
\xb \outl{Quantum Channel, Choi Matrix and Properties, How to MAP} \xa
Let $\HC_{a}, \HC_{a'}$ and $\HC_b$ be finite 
dimensional Hilbert spaces with dimensions $d_{a}=d_{a'}=d$ and $d_b$, respectively.
Let $\NC:\LC(\HC_{a'}) \mapsto \LC(\HC_b)$ be a quantum channel, and 
$\IC_a:\LC(\HC_a) \mapsto \LC(\HC_a)$ be the identity map on operators. 
Let $\{\ket{a_i}\}$ and $\{\ket{a'_i}\}$ be orthonormal basis of $\HC_{a}$ and 
$\HC_{a'}$ respectively, and $\ket{\phi} = \sum_{i} \ket{a_i}\ket{a'_i}/\sqrt{d}$ 
be a maximally entangled bipartite state.
The channel $\NC$ can be completely characterised
by a bipartite quantum state, sometimes called
the Choi matrix or the dynamical operator:
\begin{equation}
\Up = (\IC \ot \NC )\ket{\phi}\bra{\phi}, \quad \Up \in \LC(\HC_{ab}).
\label{Choi}
\end{equation}
The channel $\NC$ is completely positive if and only if the operator $\Up$ is 
positive semi-definite (\cite{Choi1975}, see \cite{PatGrif11} for a diagrammatic proof),
and $\NC$ is trace preserving if 
\begin{equation}
\Tr_{b}(\Up)= \mathbb{I}_a/d,
\label{partTr}
\end{equation}
where $\Tr_{b}$ is the partial trace over $\HC_b$, and $\mathbb{I}_a$
is the identity operator on $\HC_a$. For any $A \in \LC(\HC_{a'})$, one can show that
\begin{equation}
\NC(A) = d \, \Tr_{a}[ (A^T \ot \mathbb{I}_b )\Up],
\label{chanOut}
\end{equation}
where $A^T$ denotes the transpose of $A$
in the $\{\ket{a_i}\}$ basis. So $\NC$ is determined by the Choi matrix~$\Up$.
Equation \eqref{Choi} gives a one to one correspondence 
between $\MC_{ab}$: the convex set of quantum channels mapping $\LC(\HC_a)$
to $\LC(\HC_b)$, and $\TC_{ab}$: the convex set of density operators in 
$\LC(\HC_{ab})$ with partial trace on $\HC_b$ equaling $\mathbb{I}_a/d$.
This correspondence can be used to construct a MAP estimator 
for a quantum channel as follows. 

Suppose measurements are performed with the aim of characterising
the quantum channel $\NC$~(see \cite{DLeu03, NC97Tomo} for examples)
and data $\dl$ are collected. As in Sec. \ref{sct3.2}, let 
$p(\Up)$ be a prior probability density on $\TC_{ab}$, and 
$\Pr(\dl|\Up)$ the probability of obtaining $\dl$ given $\Up$.
The MAP estimator for the Choi matrix
\begin{equation}
 \hat{\Up}_{\text{MAP}} = \underset{\Up \in \TC_{ab}}{\text{argmax}} \; [ \log \Pr(\dl|\Up) + \log p(\Up) ],
  \label{mapEst}
\end{equation}
becomes a MAP estimate $\hat\NC_{\text{MAP}}$ for the quantum channel
when $\hat{\Up}_{\text{MAP}}$ is inserted in Eq. \eqref{chanOut}.
When the objective function 
on the right hand side of Eq. \eqref{mapEst} is concave in $\Up$, $\hat{\Up}_{\text{MAP}}$ 
can be efficiently computed using tools of convex optimization. 

\section{Modelling Noise}
\label{sct4}
\xb \outl{Noise before measurement, 
MAP estimator of state before and after noise, generality and
efficiency, comparison with other noise models} \xa
Noise is present in any experimental setup. If its effect upon a 
tomographic measurement can be modeled by assuming a known
noisy channel $\NC$~(whose parameters have been determined
by previous calibration measurements) preceding the 
final measurement as in Fig. 1, then
\begin{figure}[h]
\centering
 \includegraphics{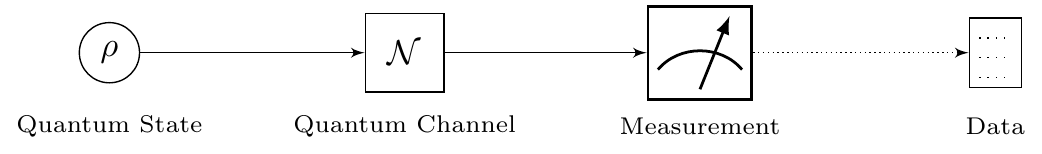}
 \caption{Schematic diagram of a quantum state $\rho$ passing through a quantum channel
		$\NC$ prior to being measured.}
\end{figure}
Bayesian estimators for $\rho$ can be obtained by replacing
$\Pr(\dl|\rho)$ with $\Pr(\dl|\NC(\rho))$ in \eqref{bayes}
and \eqref{postDen}. The MAP estimate in \eqref{MAPForm}
becomes
\begin{equation}
  \hat{\rho}_{\text{MAP}} = \underset{\rho \in \SC}{\text{argmax}} \; [ \log p(\dl|\NC(\rho)) + \log p(\rho)],
   \label{MAPFormNoise}
 \end{equation}
In addition the MAP estimate of the state coming out of the channel can be 
shown to be,
\begin{equation}
   \hat{\sigma}_{\text{MAP}} =\NC(\hat\rho_{\text{MAP}}).
   \label{MAPAfter}
\end{equation}
The above construction is quite general. There is no restriction 
on $\NC$, the form of $\Pr(\dl|\rho)$, $p(\rho)$, or the size of 
the quantum system $d$. Because $\NC$ is a linear map, if 
$\log \Pr(\dl|\rho)$ is concave in $\rho$ so is $\log \Pr(\dl|\NC(\rho))$. Thus
when the tools of convex optimization allow an efficient calculation of the MAP 
estimate in \eqref{MAPForm} the same will be true of \eqref{MAPFormNoise}.
Since the MLE, MLME, and the HMLE are special cases of MAP, they can likewise be adapted
to the the noisy setting. 

Note that the Gaussian noise models considered in \cite{Smolin2012, Singh2016} 
are quite different: they are not based on a noisy channel, but instead
on a special form of $\Pr(\dl|\rho)$.

\subsection{Example}
\label{sct4.1}
\xb \outl{compare standard MLE estimate with noisy MAP for qubits under amplitude damping
and dephasing noise} \xa
Let $\{\sg_s \}_{s \in \{x,y,z\}}$ be the Pauli matrices. A qubit density 
operator can be expressed in the Bloch parametrization,
\begin{equation}
     \rho(\rB) = \frac{1}{2}(\mathbb{I} + \rB.\vec{\sg})
     := \frac{1}{2}(\mathbb{I} + r_x \sg_x + r_y \sg_y + r_z \sg_z),
    \label{qBitBloch}
\end{equation}
where the Bloch vector $\rB = (r_x,r_y,r_z)$, with norm $|\rB|$,
belongs to the set of real variables $R=\{\rB \; | |\rB| \leq 1\}$ 
called the Bloch ball. 
For doing tomography suppose $3N$ measurements, $N$ each in the
eigenbasis $\{\ket{+}_s, \ket{-}_s\}$ of $\{\sg_s\}_{s \in \{x,y,z\}}$, are performed on identically
prepared quantum systems, each described by the density operator $\rho$. 
The probability of observing $\ket{+}_s$,
\begin{equation}
 p_s = \Tr( \rho [+]_s) = (1 + r_s)/2, \quad s \in \{x,y,z\},
\label{pDef}
 \end{equation}
where $[+]_s$ denotes the projector on $\ket{+}_s$.
The measurement data set $\dl = \{n_s, N-n_s\}_{s \in \{x,y,z\}}$, where $n_s$
denotes the number of times $\ket{+}_s$ is observed, has a probability
\begin{equation}
\Pr(\dl|\rho)  = C_{\dl} \underset{s \in \{x,y,z\}}{\prod} p_s^{n_s}(1-p_s)^{N-n_s},
\label{modDen}
\end{equation}
where $C_{\dl} = (3N)!/(\prod_s n_s!(N-n_s)!)$. 
Using \eqref{MLE} and \eqref{qBitBloch}, a simple MLE estimate 
\begin{equation}
\hat{\rho}_{\text{MLE}} = \underset{\rB \in R}{\text{argmax}} \; \log \Pr(\dl|\rho(\rB))
\label{MLECon}
\end{equation}
can be obtained. However most tomography setups have noise.
For example, when discussing noise in a nuclear magnetic 
resonance~(NMR) experiment~\cite{Vandersypen2001, PhysRevA.70.032324, PhysRevA.97.022302}
one may use the model described in Fig. 1, where the channel $\NC$, acting over time $t$ represents
the combined action of two channels, the generalized amplitude damping~($T_1$) channel,
\begin{equation}
\AC(\rho) = \sum_i A_i \rho A_i^{\dag},
\label{AD}
\end{equation}
where,
\begin{equation}
    A_0  = \sqrt{p}\begin{pmatrix} 1&0\\ 0&\sqrt{1-\gamma} \end{pmatrix},
    A_1  = \sqrt{p}\begin{pmatrix} 0&\sqrt{\gamma}\\ 0&0 \end{pmatrix},
    A_2  = \sqrt{1-p}\begin{pmatrix} \sqrt{1-\gamma}&0\\ 0&1 \end{pmatrix},
    A_3  = \sqrt{1-p}\begin{pmatrix} 0&0\\ \sqrt{\gamma}&0 \end{pmatrix},
\end{equation}
$p \in [0,1]$, is the probability of finding the qubit in the $\ket{0}$ state as 
$t \mapsto \infty$, $\gamma = 1 - e^{-t/T_1}$, $T_1$ is a time constant,
and the phase damping~($T_2$) channel,
\begin{equation}
\BC(\rho) = \sum_j B_j \rho B_j^\dag,
\label{PD}
\end{equation}
where,
\begin{equation}
    B_0  = \begin{pmatrix} 1&0\\ 0&\sqrt{1-\lm} \end{pmatrix},
    B_1  = \begin{pmatrix} 0&0\\ 0&\sqrt{\lm} \end{pmatrix},
\end{equation}
$\lm = 1 - e^{-t/T_2}$, and $T_2$ is a time constant, and
$\NC = \AC \circ \BC = \BC \circ \AC$
\footnote{
In general, a quantum channel obtained by first applying some quantum channel
$\AC$ and then some channel $\BC$ is different from the one obtained by applying
$\BC$ first and then $\AC$. However, in this special case where $\AC$ is
the qubit amplitude damping and $\BC$ is the qubit phase damping channel, 
changing the order has no effect.}.

A MAP estimate which accounts for the noise, assuming a prior probability 
density independent of $\rB$~(under this probability density function, equal volumes of the Bloch ball
have equal probability), can be obtained by using \eqref{MAPFormNoise} and \eqref{qBitBloch}, 
\begin{equation}
  \hat{\rho}_{\text{MAP}} = \underset{\rB \in R}{\text{argmax}} \; \log \Pr(\dl|\NC(\rho(\rB))).
  \label{MAPCon}
\end{equation}

We perform numerical simulations to test performance of MLE and MAP in 
\eqref{MLECon} and \eqref{MAPCon} respectively.
There is no unique metric to assess how close an estimate $\hat{\rho}$ is to the actual $\rho$. 
The fidelity $\FC(\rho,\hat{\rho}) \equiv \Tr( \sqrt{\sqrt{\rho}\hat{\rho}\sqrt{\rho}} )$
and trace distance $ \DC(\rho,\hat{\rho}) \equiv \frac{1}{2}||\rho - \hat{\rho}||_1$ are, however
popular choices.
Various parameter choices for $\AC$ and $\BC$ are used in the simulation by fixing
$p = 1/2$, $T_1/T_2 = 10$~\cite{PhysRevA.97.022302}, $t = k T_2$ and varying $k
\in \{0.25,.5,.75,1.0,1.5,2.0,2.5 \}$.
For any fixed $k$, we choose $2.5 \times 10^3$ qubit states uniformly in the Bloch ball.
For each state we simulate the construction of the MLE and MAP estimates for various $N$ values.
By averaging over all qubit states we arrive at the average log-infidelity~($\log_{10}[1 - \FC(\rho,\hat{\rho})]$) and 
log-trace distance~($\log_{10}[\DC(\rho,\hat{\rho})]$)
of the MAP and MLE estimators~(see Appendix for numerical techniques).
For a fixed $k$, under both log-infidelity and log-trace distance, the behaviour of MAP and MLE 
estimators with the number of measurements~($3N$) is shown in Fig. 2.  
\begin{figure}[h]
  \centering
  \includegraphics[scale=.55]{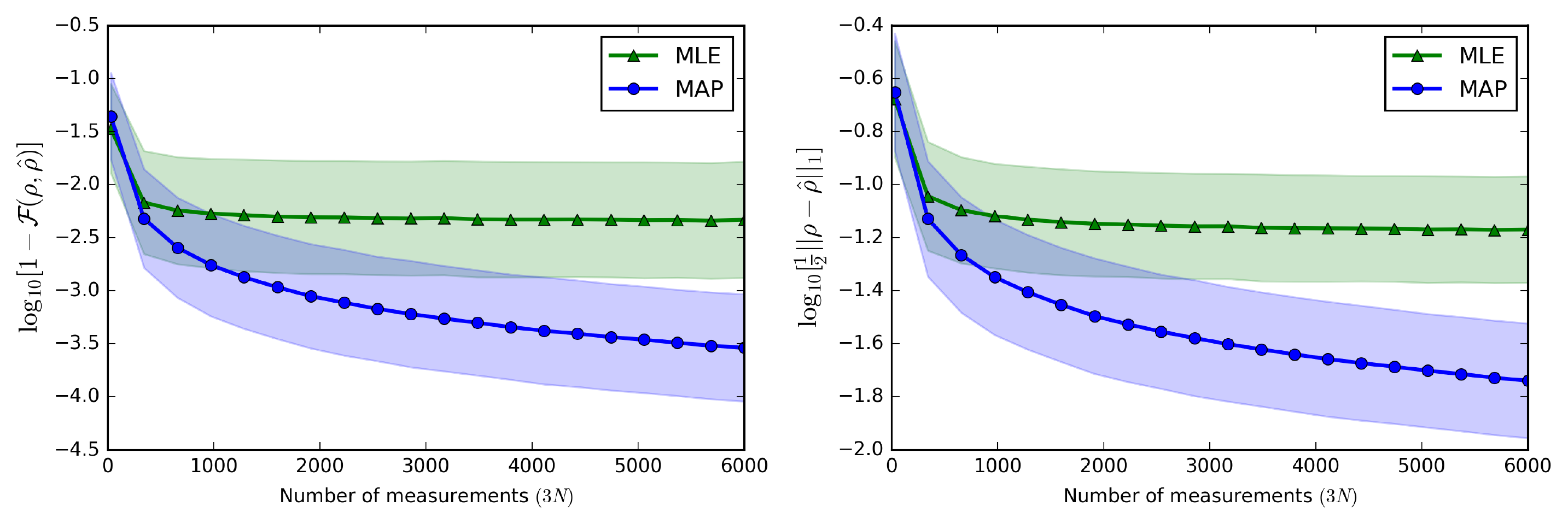}
\caption{(Color online)~Plot of the average log-infidelity and log-trace distance between the true
and estimated state, against the number of measurements at $k=0.5$. The error bars represent one standard deviation.
If the quantum channel $\NC$ was perfect and the number of measurements was infinite, the log infidelity
and log trace distance would be negative infinity for MAP and MLE. 
As the ordinate of the graph increases the performance of the estimator becomes worse.
}
\label{fig:pdfig}
\end{figure}
Under both metrics, MAP and MLE show qualitatively different behaviour.
As the number of measurements increases from a small value, for both
metrics, the average MAP value decreases and the average MLE value decreases 
but settles to a fixed number. Thus, on average the MAP estimate
always improves with the number of measurements while the MLE improves up to a point
and then ceases to change.
$\NC$ becomes more noisy as $k$ increases and this causes both the 
MAP and MLE curves in each of the plots in Fig. 2 to shift upwards. 
The upward shift implies that for a fixed $N$ with increasing $k$,
on average the MAP and MLE estimates become worse under both metrics.
In the case of MAP this effect of increasing $k$ can
be mitigated by increasing the number of measurements which on average
improves the estimate, however this is not always possible for MLE 
whose average value, under both metrics becomes fixed beyond a certain number of measurements and 
this fixed value also increases with $k$.
Under both metrics and for all $k$ values tested, 
when the number of measurements is modestly large~(roughly greater than $300$
in our case), on average MAP outperforms MLE. As the number of measurements 
increases further the average MAP value becomes an order of magnitude better 
than average MLE value, and eventually the error bars~(one standard deviation 
about the average value) on the MAP and MLE curves cease to overlap.
Thus numerical evidence on qubits shows, that except when the number
of measurements are low, on average there is an advantage of using a 
MAP estimator which accounts for noise over a standard MLE.

\section{Conclusion}
\label{sct5}
\xb \outl{MAP estimation, prior knowledge and fast numerical algorithm,
connection with other estimators, MAP with noise, numerical study and comment about 
MAP for quantum process tomography} \xa

The maximum a posteriori probability~(MAP) estimation framework 
for quantum state and process tomography
introduced here combines a number of previous quantum state
estimators, in particular the maximum likelihood, 
hedged maximum likelihood, and maximum likelihood-maximum entropy estimator,
in a single framework using Bayesian methodology.
In several cases of interest to quantum state tomography the MAP estimator
becomes a convex optimization problem which should be numerically more tractable
than the Bayes’ Mean Estimator.
Using the Choi-Jamio\l{l}kowski isomorphism, 
MAP estimation of quantum states can be extended to quantum channels,
and the extension is expected to have similar advantages as the MAP estimate for
quantum states.
When the experimental noise can be represented by a known noisy
channel preceding the measurement, the MAP estimator can be modified
to take it into account and can be computed efficiently as
long as the posterior probability density is log concave. 
Numerical results on qubits indicate that on average, such modifications
can vastly improve estimates.
Having a measure of 
reliability for any estimate is of significant value,
and it would be interesting to construct such measures for the MAP estimate.

\xb
\section*{Acknowledgments}
\xa
I am indebted to Robert B. Griffiths for valuable discussions, and thank
Renato Renner, Yong Siah Teo, Carlton Caves, Ezad Shojaee and Simon Samuroff
for their comments. This work used the Extreme Science and Engineering Discovery 
Environment (XSEDE)~\cite{XSEDE}, which is supported by National Science Foundation grant 
number ACI-1548562. Specifically, it used the Bridges system~\cite{bridges}, which is supported 
by NSF award number ACI-1445606, at the Pittsburgh Supercomputing Center (PSC).

\appendix
\section{Appendix. Convex Optimization over qubit density operators}
Projected gradient descent is a very general iterative algorithm, used
extensively for minimizing functions defined over convex sets. When the function
is convex, successive function values obtained during the algorithm approach 
the global minimum. For some special convex sets, tools from convex analysis 
can be used to compute a bound on how close a given function value is to the global minimum.
We provide an exposition of the projected gradient descent algorithm
for minimizing any differentiable convex function over the set of qubit density operators,
and illustrate a technique for checking how far a given function value is from the 
global minimum. 
Let $f:R \mapsto \mathbb{R}$ be a differentiable convex function, then
\begin{equation}
f(\xB^*) = f^* = \underset{\xB \in R}{\min} \; f(\xB),    
\end{equation}
is called the optimization problem, with objective function $f$,
optimal $\xB^*$ and optimum value $f^*$.
Projected gradient descent begins with some point inside $R$, 
then iteratively takes steps to move to a new point with a lower function value. 
The algorithm halts when some \textit{stopping criterion} is met. 
New points are chosen by moving along the negative gradient direction by an amount 
called the \textit{step size}, such movements may take one outside the set $R$, 
in which case we project onto the boundary of the set. If $\xB$ is a vector in 
$\mathbb{R}^3$ then its projection onto $R$,
\begin{equation}
P_{R}(\xB) =  \begin{cases}
\xB/|\xB|     \quad \text{if} \quad |\xB| > 1 \\
\xB \quad \text{otherwise}
              \end{cases}.
\end{equation}
A pseudo code for projected gradient descent is given in algorithm 1.

\RestyleAlgo{boxruled}
\begin{algorithm}[H]
 \DontPrintSemicolon
 \SetAlgoLined
 \SetKwInOut{Input}{Input}\SetKwInOut{Output}{Output}
 \Input{$\xB \in R$}
 \Output{$\tilde{\xB}$ and $\tilde{f}$}
\BlankLine
\While{not stopping criterion}{
$\dB = \nabla f(\xB)$ \;
$\yB_t = \xB - t \dB$ \;
$\xB^+(t) = P_{R}(\yB(t))$ \;\label{alg1:proj}
Choose step size $t^*$ \;\label{alg1:step}
$\xB = \xB^+(t^*)$ }
\Return{$\xB, f(\xB)$}\;
\caption{Projected gradient descent}
\end{algorithm}
There are several different ways of choosing a step size and a stopping criterion.
We select the step size by using a method called \textit{backtracking line search}.
Let
\begin{equation}
G_t(\xB) \equiv (\xB - \xB^+(t))/t 
\label{genGrad}
\end{equation}
be the generalized gradient at $\xB$ for step size $t$. A pseudo code for
backtracking, to be used as a subroutine in algorithm 1 is provided in algorithm 2.

\RestyleAlgo{boxruled}
 \begin{algorithm}[H]
 \DontPrintSemicolon
 \SetAlgoLined
 \SetKwInOut{Input}{Input}\SetKwInOut{Output}{Output}
 \Input{$\beta \in (0, 1)$}
 \Output{$t^*$}
\BlankLine
Let $t = 1$\;
\While{$f(\xB^+(t)) > f(\xB) - t \inpV{\dB}{G_t(\xB)}  +  \frac{t}{2} |G_t(\xB) |^2$}{
 $t = \beta t$ \; }
 \Return{$t$}\;
 \caption{Backtracking line search}
\end{algorithm}
Backtracking line search ensures that the function
decreases by at least $t^*|G_{t^*}(\xB)|^2/2$ in each iteration~\cite{armijo1966}.
One possible stopping criterion is to check whether the function
hasn't decreased appreciably over the past few iterations or 
$|G_{t^*}(\xB)|$ is greater than a small constant. 
At any point $\xB \in R$, the surrogate duality gap~\cite{jaggi13}
\begin{equation}
      g(\xB) = \underset{\rB \in R}{\max} \; \langle \nabla f (\xB), \xB - \rB \rangle
\end{equation}
upper bounds $|f(\xB) - f^*|$ and provides a measure of how close a function value 
at $\xB$ is to the global minimum value. Due to the simple structure of the set of qubit density operators,
the surrogate duality gap can be easily computed. Let $\nabla F(\xB) = \nabla f (\xB).\vec{\sg}$
be a matrix, $\lm_{\min}(\nabla F(\xB))$ be its smallest eigenvalue, then 
\begin{equation}
    g(\xB) = \langle \nabla f (\xB), \xB \rangle - \lm_{\min}(\nabla F(\xB)).
\end{equation}
Given the ease of computing $g(\xB)$, 
checking whether its value is greater that a small constant also serves
as a good stopping criterion. Once algorithm 1 converges and returns some 
$(\tilde{\xB}, f(\tilde{\xB}))$, computing $g(\tilde{\xB})$ gives an estimate of 
how far $f(\tilde{\xB})$ is from $f^*$.

The projected gradient algorithm discussed above can be used to solve optimization problems in Eq.
\eqref{MLECon} and \eqref{MAPCon} to obtain $\hat{\rho}_{\text{MLE}}$ and $\hat{\rho}_{\text{MAP}}$ respectively.
The gradient of the objective function in these optimization problems can be computed analytically.
Let $\rB = (r_x, r_y, r_z)$ be a Bloch vector for a qubit density 
operator $\rho$, then upto local unitaries at the input and output of a qubit channel 
$\NC$, the Bloch vector $\rB'$ for $\NC(\rho)$ can always be written as 
$\rB' = (l_x r_x + t_x, l_y r_y + t_y, l_z r_z + t_z)$~\cite{GriffithsNotesChan}. 
When $\NC = \IC$
\begin{equation}
l_s = 1, t_s = 0, \; s \in \{x,y,z\},    
\end{equation}
when $\NC = \AC \circ \BC$~(see Eq. \eqref{AD} and \eqref{PD})
\begin{equation}
    l_x = l_y = \sqrt{(1-\lm)(1-\gamma)}, \; l_z = 1-\gamma, t_x=t_y=0,
    \;\text{and}\;t_z = \gamma(2p-1).
\end{equation}
The gradient of the function on the right hand side in Eq. \eqref{MAPCon}
can be obtained using
\begin{equation}
    \frac{\partial }{ \partial r_s}\log \Pr(\dl|\NC(\rho(\rB))) 
    = \frac{l_s}{2} \big(\frac{n_s}{p_s} - \frac{N-n_s}{1-p_s} \big),
    \quad s \in \{x,y,z\}.
\end{equation}
Choosing $\NC =\IC$ gives the gradient of
the function in the right side of Eq. \eqref{MLECon}. 
A python implementation of projected gradient descent
and surrogate duality gap computation over qubits,
including a code that simulates the tomography reconstruction
and generates plots for Fig. 2 
is publicly available~\cite{MAPQubit}.

\bibliographystyle{unsrt}

\end{document}